\documentclass[12pt]{article}
\usepackage{cyr}
\usepackage{epsf}
\usepackage{graphics}
\usepackage{color}
\usepackage{pazha}
\tightenlines

\def\*{$^{*}$}
\def\a{$^{\mbox{\tiny a}}$}
\def\b{$^{\mbox{\tiny b}}$}
\def\c{$^{\mbox{\tiny c}}$}
\def\d{$^{\mbox{\tiny d}}$}
\def\e{$^{\mbox{\tiny e}}$}
\def\f{$^{\mbox{\tiny f~}}$}
\def\g{$^{\mbox{\tiny g}}$}
\def\h{$^{\mbox{\tiny h}}$}
\sloppypar

\begin{document}
\begin{flushleft}
{\it to be published in Astronomy Letters, 2017, v. 43, n. 7,
  pp. 464--470}\\ [30mm]
\end{flushleft}

\baselineskip 21pt

\title{\bf THE ORIGIN OF THE BIMODAL
  LUMINOSITY DISTRIBUTION OF ULTRALUMINOUS X-RAY PULSARS}

\author{\bf S.A.~Grebenev\affilmark{*}}   

\affil{{\it Space Research Institute, Russian Academy of
    Sciences, \\ Profsoyuznaya ul. 84/32, Moscow, 117997 Russia}}

\vspace{2mm}
\received{November 14, 2016}
\sloppypar 
\vspace{2mm}
\noindent
The mechanism that can be responsible for the bimodal luminosity
distribution of super-Eddington X-ray pulsars in binary systems
is pointed out. The transition from the high to low state of
these objects is explained by accretion flow spherization due to
the radiation pressure at certain (high) accretion rates. The
transition between the states can be associated with a gradual
change in the accretion rate. The complex behavior of the
recently discovered ultraluminous X-ray pulsars M 82 X-2, NGC
5907 \mbox{ULX-1}, and NGC 7793 P13 is explained by the proposed
mechanism. The proposed model also naturally explains the
measured spinup of the neutron star in these pulsars, which is
slower than the expected one by several times.

\noindent
{\bf DOI:} 10.1134/S1063773717050012


\vfill
\noindent\rule{8cm}{1pt}\\
{$^*$ E-mail $<$sergei@hea.iki.rssi.ru$>$}

\clearpage
\section*{INTRODUCTION}
\noindent 
The discovery (Bachetti et al. 2014) of X-ray pulsations with a
mean period $P_s\simeq 1.37$~s from the ultraluminous X-ray
(ULX) source \mbox{M\,82\,X-2} (=NuSTAR\,J095551+6940.8) and its
sinusoidal modulation with a period $P_b\simeq2.5$ days (the
orbital period of the binary system) changed drastically our
views of the nature of ULX sources. Previously, it had been
assumed that a high observed isotropic X-ray luminosity of such
sources $L_{\rm iso}\ga 10^{40}\ \mbox{erg s}^{-1}$ could be
reached only during accretion onto a black hole with a
moderately large, $\sim10^3\ M_{\odot}$, or at least stellar,
$\sim10\ M_{\odot}$, mass (provided the formation of a
relativistic jet and associated strong radiation anisotropy). It
has now become clear that such a luminosity can also take place
during accretion onto a neutron star possessing a strong
magnetic field with a mass of only $M_*\sim1.4\ M_{\odot}$. Such
binary systems must be widespread and can even dominate in the
population of ULX sources (Shao and Li 2015). The discoveries of
the ultraluminous X-ray pulsars NGC\,7793 P13 and NGC\,5907 ULX-1
by the XMM-Newton satellite (Israel et al. 2017a, 2017b) shortly
afterward confirm this point of view and give hope for the
detection of other objects of this type. Note that NGC\,5907
ULX-1 has a record peak luminosity even for ULX sources, in
particular, it exceeds the maximum detected luminosity of
\mbox{M\,82\,X-2} by several times (see the table). 

The discovery of ULX pulsars has thrown down a serious challenge
to theorists. For example, it is still unclear, though is widely
discussed, how such a high luminosity is reached, which exceeds
the Eddington one for spherically symmetric accretion onto a
neutron star by hundreds of times:
\begin{equation}\label{led}
L_{\rm ed}=\frac{4\pi GM_*m_p c}{\sigma_{\rm es}}\simeq
1.9\times 10^{38} \left(\frac{\sigma_{\rm T}}{\sigma_{\rm es}}\right)
\left(\frac{M_*}{1.4\ M_{\odot}}\right)\ \mbox{erg s}^{-1}.
\end{equation}

Here, $\sigma_{\rm es}$ is the electron scattering cross
section, $\sigma_{\rm T}$ is the Thomson cross section, $G$ is
the gravitational constant, $m_p$ is the proton mass, and $c$ is
the speed of light. Of course, the accretion onto a neutron star
with a strong magnetic field is far from spherically symmetric
one. As early as 1976, having considered a realistic accretion
flow geometry at a supercritical accretion rate, Basko and
Sunyaev (1976) showed that the isotropic luminosity of a pulsar
could exceed $L_{\rm ed}$ by more than an order of magnitude
(see below).  Nevertheless, it is still insufficient to explain
the observations of ULX pulsars.

Many of the authors (e.g., Lyutikov 2014; Tong 2015; Eksi et
al. 2015; Tsygankov et al. 2016a; Israel et al. 2017a, 2017b)
are inclined to the assumption about an extreme magnetic field
strength of the neutron star in ULX systems ($B_*\ga 10^{14}$
G), which reduces the electron scattering cross section
$\sigma_{\rm es}$ and, thus, raises the Eddington limit. Others
(e.g., Kluzniak and Lasota 2015) think that a high luminosity is
reached precisely because of the reduced (to $B_*\sim10^{9}$ G)
magnetic field strength (compared to its values
$B_*\sim10^{12}-10^{13}$~G typical for X-ray pulsars).  Because
of the weak magnetic field, the accretion disk almost reaches
the neutron star surface and radiates in the same way as during
super-Eddington accretion onto a black hole. In both cases, the
limiting observed luminosity of ULX pulsars, $L_{\rm iso}\sim
10^{41}\ \mbox{erg s}^{-1}$, still cannot be explained and one
has to appeal to a strong anisotropy of their radiation
(dall'Osso et al. 2015; Chen 2017).
\begin{table}[t]
\small
\centering {{\bf Table.} Parameters of the ULX pulsars discovered to date, their corotation, magnetospheric,\\ and spherization radii\\}

\vspace{2mm}
\footnotesize
\hspace{-5mm}\begin{tabular}{@{\,}l@{\,}|c|c|c|c|c|c|c|c|c|c|c|c@{\,}} \hline\hline
 & \multicolumn{8}{c|}{General}
 & \multicolumn{2}{c|}{In high state} 
 & \multicolumn{2}{c}{In low state}\\ \cline{2-13}
   Source& $P_{b}$\a & $P_{s}$\b &$\dot{P}_{-10}$\b& $\gamma$\c&$\mu_3$\d&$R_{\rm
     c}$\e& $\dot{M}_{20}$\f& $R_{\rm   s}$\e& $L_{39}$\g& $R_{\rm m}$\e& 
  $L_{39}$\g& $R_{\rm ms}$\h\\ 
&days & s & s s$^{-1}$ &&& km&\,g\,s$^{-1}$&km&erg\,s$^{-1}$&km&
   erg\,s$^{-1}$&km \\\hline
M\,82\,X-2             &2.5 &1.37 & -2.0&4& 3& 2080&0.59&  860&37  & 900&0.28&890\\  
NGC\,5907\,ULX-1&5.3 &1.13 & -8.1&6&12&1830&1.06 &1550&100&1670&<0.3&1640\\
NGC\,7793\,P13   &      &0.42  & -0.4&2&1.4& 950&0.41&   610&13  &640&0.3&630\\ \hline
\multicolumn{13}{l}{}\\ [-3mm]
\multicolumn{13}{l}{\a\ The orbital period $P_{b}$.}\\
\multicolumn{13}{l}{\b\ The pulsar period $P_s$ and mean period derivative $\dot{P}_s=10^{-10}\dot{P}_{-10}$. }\\
\multicolumn{13}{l}{\c\ The presumed factor of the emission anisotropy.}\\
\multicolumn{13}{l}{\d\ The presumed magnetic momentum of the neutron star $\mu=3 \times 10^{30}\ \mu_3\ \mbox{G cm}^{3}$.}\\
\multicolumn{13}{l}{\e\ The corotation, $R_{\rm c}$,
  spherization, $R_{\rm s}$, and magnetospheric, $R_{\rm m}$, radii.}\\
\multicolumn{13}{l}{\f\ The presumed accretion rate
  $\dot{M}_{0}=10^{20}\dot{M}_{20}$ corresponding to the luminosity $L_{39}$ in the high state}\\
\multicolumn{13}{l}{\ \ \ \ (according to $L^{\rm obs}_{\rm iso}=\gamma\ GM_*\dot{M}_0/R_*$).}\\
\multicolumn{13}{l}{\g\ The observed isotropic X-ray luminosity $L^{\rm obs}_{\rm
      iso}=10^{39} L_{39}$ in the energy range 0.3--10 keV.}\\ 
\multicolumn{13}{l}{\h\ The magnetospheric radius in the  low state according to Eq. (\ref{rms}).}\\ 
\end{tabular}
\end{table}

\indent The nature of the bimodal luminosity distribution of ULX
pulsars pointed out by Tsygankov et al. (2016a) and Israel et
al. (2017a, 2017b) also remains a puzzle. In addition to the
state with a very high X-ray luminosity (hereafter the high
state), periods during which the luminosity dropped to $\la
3\times10^{38}\ \mbox{erg s}^{-1}$ (hereafter the low state)
have been detected for all three sources (see the table).
Tsygankov et al. (2016a) and Israel et al. (2017а) assumed the
bimodality of the luminosity distribution to be associated with
the action of centrifugal forces, which inhibit accretion and
are capable of expelling an excess of accreting matter from the
system (the propeller effect; Illarionov and Sunyaev 1975; see
also Corbet 1996). This effect begins to manifest itself as soon
as the magnetospheric radius of the neutron star $R_m$ during
the evolution of the system (for example, a temporary decrease
in the accretion rate) exceeds the corotation radius $R_c$
(otherwise the surface rotation velocity of the magnetosphere
will exceed the Keplerian velocity). In this case, the accretion
onto the neutron star ceases, and only the radiation from the
outer disk region $R>R_m$ is observed. In order for the
propeller effect to operate in the systems being discussed, it
is necessary that the neutron stars in them possess a very
strong magnetic field $B_*\sim10^{14}-10^{15}$~G similar to the
field of magnetars (Tsygankov et al. 2016a).

Although the very existence of the propeller effect is beyond
doubt and has come into wide use by astrophysicists, i.e., it is
used to explain the observed luminosity jumps in millisecond
(LMXBs, Campana et al. 2008, 2014) and ordinary (HMXBs, Corbet
et al. 1996; Campana et al. 2002; Tsygankov et al. 2016b;
Postnov et al. 2017) X-ray pulsars, the existence of
``equilibrium'' pulsar periods (van den Heuvel 1984; Corbet
1986), the outbursts of fast X-ray transients (Grebenev and
Sunyaev 2007; Grebenev 2009), and many other observed phenomena,
the action of this mechanism as a cause of the bimodal
luminosity distribution of ULX pulsars raises doubts.  This is
not only due to the very strong neutron star magnetic field,
$B_*\ga10^{14}$~G, required for this purpose, but also due to
the observed range of the luminosity drop, which is smaller by
several times than the expected one $\sim R_c/R_*\simeq
140\ m_*^{1/3}p_*^{2/3}R_{12}^{-1}$ (Corbet 1996; Tsygankov et
al. 2016a; here, $p_*$ is the spin period of the neutron star
$P_s$ in seconds, while $m_*$ and $R_{12}$ are its mass $M_*$
and radius $R_*$ normalized to their standard values of
$1.4\ M_{\odot}$ и $12$ km), and, most importantly, the very
close coincidence of the luminosity of the sources in their low
state with the Eddington one $L_{\rm ed}$.  In this paper we
will show that there exists a different explanation for the
abrupt change of the luminosity in these sources associated with
the transitions between two different regimes of supercritical
accretion onto a neutron star with a strong magnetic field. The
transitions are caused by accretion flow spherization in the
disk due to the radiation pressure when a certain accretion rate
dependent on the magnetic field strength of the neutron star is
exceeded.

\section*{THE REGIMES OF SUPERCRITICAL ACCRETION}
\noindent
The properties of the accretion flow onto a neutron
star and the interaction of this flow with its magnetic
field are defined by four characteristic radii:
 
The {\em magnetospheric radius\/}
  \begin{equation}\label{rmag}
    R_m\simeq\xi\left(\frac{\mu_*^2}{\sqrt{2GM_*}\dot{M}}\right)^{2/7}\simeq
8.2\times 10^7 \  \xi\ \mu_{3}^{4/7} m_*^{-1/7} \dot{m}_{20}^{-2/7}\ \mbox{cm},
  \end{equation}
at which the pressure of the matter inflowing through
the accretion disk is equal to the pressure of the neutron
star magnetic field (Davidson and Ostriker 1973;
Illarionov and Sunyaev 1975);

 the {\em spherization radius\/} of the accretion flow
\begin{equation}\label{rspher}
R_s=\frac{3}{8\pi}\frac{\dot{M}_0}{m_p}\frac{\sigma_{\rm T}}{c}=
\frac{3}{2}\frac{GM_*\dot{M}_0}{L_{\rm ed}}\simeq {1.5\times 10^8}\ \dot{m}_{20}\ \mbox{cm},
\end{equation}
at which the accretion disk under radiation pressure swells so
that its half-thickness is equal to the radius $R$ (Shakura and
Sunyaev 1973)\footnote{The disk luminosity in the region $R>R_s$
  turns out then to be equal to the Eddington luminosity
  $L_d=(3/2) G M_* \dot{M}_0/R_s=L_{\rm ed}$ (Lipunova 1999).};

 the {\em corotation radius\/}
\begin{equation}\label{rcor}
   R_c=\left(\frac{GM_*P_s^2}{4\pi^2}\right)^{1/3}=1.7\times
10^8\  m_*^{1/3} p_*^{2/3}\ \mbox{cm},
\end{equation}
at which the surface rotation velocity of the magnetosphere
$(2\pi/P_s) R_c$ is equal to the Keplerian velocity
$(GM_*/R_c)^{1/2}$ (Illarionov and Sunyaev 1975);

and, of course, 
the {\em intrinsic radius\/} of the neutron star $R_*$.

\noindent
Here, $\xi\simeq0.5$ is the correction that takes into account
the deviation of the magnetospheric radius in the case of disk
accretion from the Alfv\'en radius computed  for
spherically symmetric accretion (Ghosh and Lamb 1978),
$\dot{m}_{20}$ is the accretion rate $\dot{M}_0$ in units of
$10^{20} \ \mbox{g s}^{-1}$ ($= 1.6\times
10^{-6}\ M_{\odot}\ \mbox{yr}^{-1}$), $\mu_*=0.5 B_* R_*^3$ is
the dipole magnetic moment of the neutron star, and $B_*$ is the
magnetic field strength at its poles.  The magnetic moment
$\mu_*$ expressed in units of $3\times10^{30}\ \mbox{G cm}^3$
will be denoted by $\mu_{3}$. Note that by $\dot{M}$ in
Eq.~(\ref{rmag}) we mean the accretion rate near the
magnetospheric boundary. During super-Eddington accretion in the
inner disk regions $\dot{M}$ can decrease compared to the
external value $\dot{M}_0$ due to the outflow of matter.
\begin{figure}[t]
\hspace{0.09\textwidth}\epsfxsize=0.9\textwidth
\epsffile{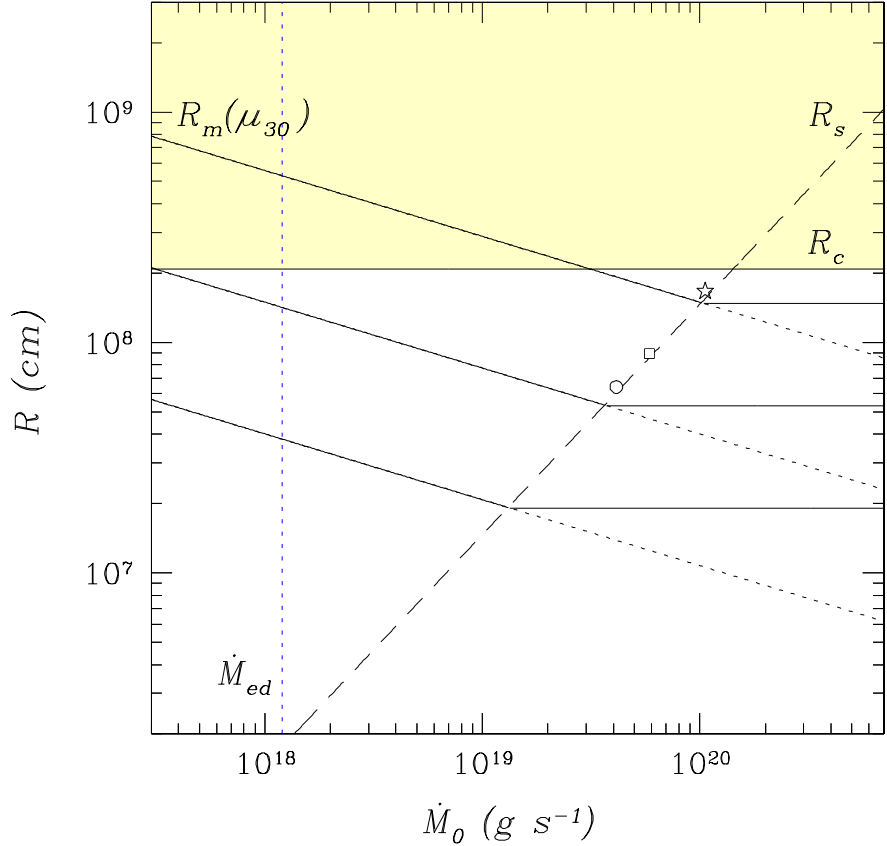}
\caption{\rm Spherization radius of the accretion flow $R_s$
  (dashed line), corotation radius $R_c$ (the lower boundary of
  the shaded region), and magnetospheric radius of the neutron
  star $R_m$ (solid lines) versus accretion rate for an
  ultraluminous X-ray pulsar with the same period as that for
  M\,82\,X-2. Different magnetic moments of the star are
  considered, $\mu_{3}=0.1,\ 1,$ and $10$ (the solid lines from
  the bottom upward). The shaded region corresponds to the
  radius $R_m$ at which the propeller regime is realized. The
  vertical dotted line marks the Eddington accretion
  rate. Asterisk, square and circle show values of $R_m$ and
  $\dot{M}_0$ adopted in the paper for the maximum high state of
  the ULX pulsars NGC 5907 ULX-1, M82 X-2, and NGC 7793 P13, respectively.}
\end{figure}

In Fig.\,1 the radii $R_m$, $R_s$, and $R_c$ are plotted against
the accretion rate for three magnetic moments of the neutron
star, $\mu_3= 0.1,\ 1,$ and $10$. These values correspond to
magnetic field strengths at the stellar poles $B_*\simeq
3.5\times10^{11}$, $3.5\times10^{12}$ и $3.5\times 10^{13}$ G,
respectively. The pulsation period was assumed to be $1.37$ s,
the same as that for the ULX pulsar M\,82\,X-2. The estimates of
the radii $R_m$, $R_s$, $R_c$ for this and the two other ULX
pulsars known to date are given in the table. As will be shown
below, once $R_s$ has reached $R_m$, the magnetospheric radius
$R_m$ ceases to depend on $\dot{M}_0$. Therefore, the dependence
(\ref{rmag}) in this region is indicated in Fig.\,1 by the
dotted line.

Figure\,1 suggests that the dependence of the magnetospheric
radius of the neutron star $R_m$ on $\dot{M}_0$ has two singular
points. This radius is equal to the corotation radius $R_c$ and
the spherization radius $R_s$ at the first and second points,
respectively. The first event occurs at an accretion rate
\begin{equation}\label{mdmc}
  \dot{M}_{mc}\simeq7.2\times 10^{17}\ \mu_{3}^{2}\, m_*^{-5/3} p_*^{-7/3}\
  \mbox{g s}^{-1},
\end{equation} 
and the second one occurs at an accretion rate (Lipunov 1982)
\begin{equation}\label{mdms}
  \dot{M}_{ms}\simeq 3.7\times 10^{19}\ \mu_{3}^{4/9} m_*^{-1/9}\
  \mbox{g s}^{-1}.
\end{equation} 
At $\dot{M}_0\le\dot{M}_{mc}$ no efficient accretion is possible
because of the propeller effect --- the infalling matter is ejected
from the system. At $\dot{M}_0\ge\dot{M}_{mc}$ and up to
$\dot{M}_0\simeq\dot{M}_{ms}$ nothing inhibits it; the regime of
direct accretion observed in ordinary X-ray pulsars with the
modifications for $\dot{M}_0 \ga \dot{M}_{\rm ed}=L_{\rm
  ed}R_*/(GM_*)\simeq1.2\times10^{18}\ \mbox{g s}^{-1},$
described by Basko and Sunyaev (1976), is realized. Note that in
this regime the spherization radius $R_s<R_m$. Let us consider
the case of direct supercritical accretion in more detail.
 
\subsection*{The High State (an Accretion Rate
  $\dot{M}_{mc}\leq\dot{M}_0\leq\dot{M}_{ms}$)}
\noindent
In this case, just as in the case of ordinary X-ray pulsars,
upon reaching the boundary of the magnetosphere, the accretion
disk matter is frozen into its upper layer and is transferred by
two streams to high-latitude regions, where it flows down along
the so-called accretion columns into the vicinity of the neutron
star magnetic poles. Basko and Sunyaev (1976) (see also
Lyubarskii and Sunyaev 1988; Mushtukov et al. 2015) showed that
at $\dot{M}_0\ga\dot{M}_{\rm ed}$ the radiation flux escaping
through the walls of the accretion columns in X-ray pulsars
exceeds the radial (Eddington) flux by a factor of $\sim
H/d\simeq 240\ R_{12}\, (H/R_*);$ accordingly, their total
luminosity
$$ L_{\rm iso}=4Hl \left(\frac{H}{d}\right) \left(\frac{L_{\rm ed}}{4\pi R_*^2}\right)= 
\frac{l}{\pi d}\left(\frac{H}{R_*}\right)^2L_{\rm ed}.
$$ can exceed $L_{\rm ed}$. Here, $H$ is the height of the accretion
columns\footnote{To be more precise, the height of the radiation-dominated
shock in which the matter sinking in the column walls is
heated to high temperatures above the neutron star surface.},
$l\simeq2.5\times10^{5}$ cm is the width of their base,
$d\simeq5\times10^3$ cm is the thickness of the walls. Since
typically $H\la R_*$, the actual increase in luminosity is
limited by
\begin{equation} L^{\rm max}_{\rm iso}\la 3\times10^{39}
\left(\frac{l/d}{50}\right)
\left(\frac{\sigma_{\rm T}}{\sigma_{\rm es}}\right)
\left(\frac{M_*}{1.4\ M_{\odot}}\right)\ \mbox{erg s}^{-1}.\label{lbasko}
\end{equation} The luminosity of the accretion disk $L_{\rm
  d}\la L_{\rm ed}$ should be added to the luminosity of the
accretion columns $L_{\rm iso}$, but still, to achieve agreement
with the observations of ULX pulsars, it is necessary either to
assume an appreciable radiation anisotropy or to take into
account the decrease in the scattering cross section due to a
strong magnetic field (Basko and Sunyaev 1975, 1976). Indeed, in
the presence of a magnetic field at energies
$E<E_{B}=11.6\,(B_*/10^{12}\ \mbox{G})$ keV the electron
scattering cross section $\sigma_{es}$ decreases compared to the
Thomson one as $\sigma_{\rm X}\simeq\sigma_{\rm T}(E/E_{B})^2$
for the extraordinary wave and as $\sigma_{\rm
  O}\simeq\sigma_{\rm T}[\sin^2\theta+(E/E_{B})^2]$ for the
ordinary one; here, $\theta$ is the angle between the direction
of wave propagation and the magnetic field lines. Since the
emergent radiation is multiply scattered in the accretion column
walls, with the ordinary and extraordinary waves being
transformed into one another, the effective scattering cross
section in the standard X-ray band ($E\la10$ keV) can be
appreciably smaller than $\sigma_{\rm T}$ (Paczynski
1992). Introducing an anisotropy factor $\gamma>1$, suggesting
that the radiation intensity toward us is greater than the mean
intensity by a factor of $\gamma$, from inequality
(\ref{lbasko}) we finally obtain
\begin{equation} L^{\rm max}_{\rm aniso}\la 3\times
10^{40}\ \left(\frac{\gamma\, \sigma_{\rm T}/\sigma_{\rm
  es}}{10}\right)\ m_*\ \mbox{\rm erg s}^{-1}.\label{lbasko2}
\end{equation}
The parameter $\epsilon=\gamma\, (\sigma_{\rm T}/\sigma_{\rm es}),$
which we set equal
to $10,$ characterizes the joint uncertainty in the
anisotropy of the emergent radiation and the decrease
in the scattering cross section. The pulse profile
for ULX pulsars is fairly smooth, nearly sinusoidal
(Bachetti et al. 2014; Israel et al. 2017a, 2017b).
Given that it is shaped by the radiation emerging from
the walls of the accretion columns, it is hard to expect
a very strong anisotropy of this radiation. Below we
assume that $\gamma=2-4.$

If the accretion occurred with the maximum possible
efficiency, then one would expect the observed luminosity to be
\begin{equation}\label{lobs}
L^{\rm obs}_{\rm iso}=\gamma\,GM_*\dot{M}_0/R_*\simeq
1.6\times10^{40}\ \gamma\,R_{12}^{-1} 
m_*\dot{m}_{20}\ \mbox{erg s}^{-1}.
\end{equation}
Comparing this expression with inequality (\ref{lbasko2}), we
see that the energy being released during accretion can be
efficiently reprocessed into radiation only as long as
$\dot{m}_{20}\leq 0.2 R_{12} (\sigma_{\rm
  T}/\sigma_{es})$. As the accretion rate increases further, no
rise in luminosity occurs, the excess of energy being released
is carried away to the neutron star surface (Basko and Sunyaev
1976).  Note that the adopted values of $\epsilon$ and $\gamma$
allow the observed maximum luminosities of the ULX pulsars
M\,82\,X-2 and NGC\,7793 P13 to be explained (see the table). In
the case of NGC\,5907 ULX-1, however, $\epsilon=\gamma
(\sigma_{\rm T}/\sigma_{\rm es})$ and especially $\gamma$ should
be additionally increased by a factor of 2--3. A much stronger
magnetic field apparently operates in this source, which leads
to a more noticeable decrease in the scattering cross section
$\sigma_{\rm es}$, a more significant increase in the Eddington
limit, and a more strong anisotropy of the radiation.

\subsection*{The Low State (a High Accretion Rate $\dot{M}_0\geq\dot{M}_{ms}$)}
\noindent
The spherization radius is equal to $R_m$ at an accretion
rate $\dot{M}_0\simeq\dot{M}_{ms}$ and begins to exceed it as 
$\dot{M}_0$ increases further. In this case: (1) the accretion disk
swells near $R_s$, (2) an efficient outflow of excess matter
and angular momentum with a nearly parabolic
velocity is formed above the disk at $R<R_s$, and
(3) the accretion in the region $R_m<R<R_s$ occurs in a
regime close to the spherically symmetric one with a
rate decreasing as
\begin{equation}\label{ss-sol}
\dot{M}(R<R_s)=\dot{M}_0\ R/R_s
\end{equation}
(Shakura and Sunyaev 1973; Lipunova 1999). The total luminosity
of the source in this case does not exceed $\simeq 2\,L_{\rm
  ed}.$ Half of it, $\simeq L_{\rm ed}$, is emitted by the outer
$R>R_s$ disk regions, and the other half is emitted by the inner
envelope formed by the outflowing matter.  Irrespective of
precisely where and how the energy release occurs here, due to
the quasi-sphericity of this envelope, the luminosity of the
radiation leaving it cannot exceed $\simeq L_{\rm ed},$ the
remaining energy being released is spent on the acceleration of
the outflowing matter\footnote{Note that even formally the above
  solution (\ref{ss-sol}) for the decrease in the accretion rate
  at $R<R_s$ was obtained by assuming that the {\sl entire\/}
  energy being released at $R<R_s$ is spent on the radiation
  acceleration of the outflowing matter (Lipunova
  1999). Therefore, the frequently encountered assertion that
  the inner region gives a logarithmic $\sim L_{\rm
    ed}\ln{(\dot{M}_0/\dot{M}_{\rm ed})}$ increase of the
  luminosity of the outer disk is incorrect.}. In this sense,
the observed picture has much in common with the photospheric
expansion of the neutron star atmosphere during super-Eddington
X-ray bursts (see, e.g., Lewin et al. 1993). Just as for bursts,
depending on the accretion rate, the density of the outflowing
matter and the size of its photosphere (inner envelope) and,
accordingly, the effective temperature of the emergent radiation
change.

Although, on the whole, the X-ray observations
of ULX pulsars in their low state are consistent with $\simeq 2\,L_{\rm
  ed},$ at high accretion rates an increasingly large
fraction of the radiation must fall into the ultraviolet
and optical spectral ranges. Therefore, the X-ray
luminosity in the low state for some of the sources can
be appreciably below the Eddington level. This may
be true for NGC\,5907 ULX-1 (Israel et al. 2017a).

Because of the decrease in the accretion rate at $R<R_s$, the
flux of matter reaching the boundary of the neutron star
magnetosphere turns out to be equal 
only to $\dot{M}_0 R_m/R_s$. Accordingly, the magnetospheric
radius $R_m$ does not decrease with increasing $\dot{M}_0$
after reaching the critical accretion rate $\dot{M}_{ms}$ (as
$\dot{M}_0^{-2/7}$, see Eq.~\ref{rmag}), but remains equal to
its value at $\dot{M}_0=\dot{M}_{ms},$ 
\begin{equation}
R_m^{\rm low}\simeq 5.4\times 10^{7}\ \mu_{3}^{4/9}
m_*^{-1/9}\ \mbox{cm}.\label{rms}
\end{equation}

In Fig.\,1 this part of the dependence of $R_m$ on $\dot{M}_0$
is indicated by the solid horizontal line, while the
dependence~(\ref{rmag}) is indicated by the dashed line.
\begin{figure}[t]
  \hspace{0.09\textwidth}\epsfxsize=0.9\textwidth
\epsffile{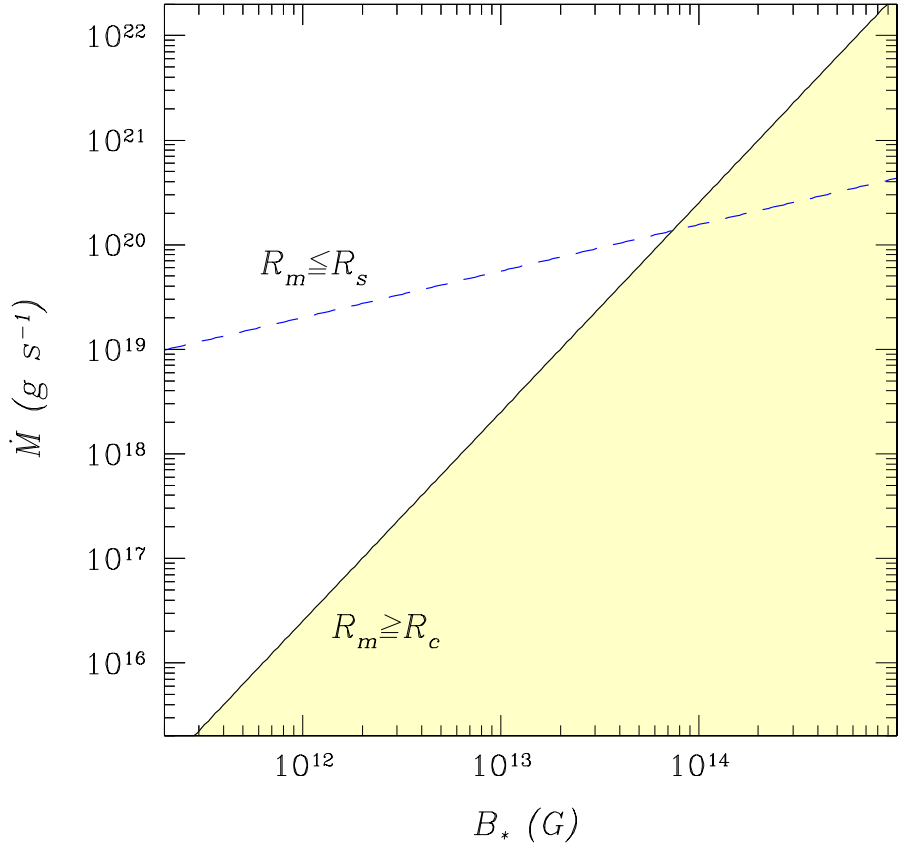}
\caption{\rm The accretion rate $\dot{M}_{mc}$ at which the
  magnetospheric radius $R_m$ is equal to the corotation radius
  $R_c$ (solid line) and the accretion rate $\dot{M}_{ms}$ at
  which $R_m$ is equal to the spherization radius $R_s$ (dashed
  line) as functions of the neutron star magnetic field strength
  $B_*$.  At $R_m>R_c$ (shaded) the direct accretion stops due
  to the propeller effect. At $R_m<R_s$ (above the dashed curve)
  the luminosity of the source is restricted by the Eddington
  limit, $L\la 2 L_{\rm ed}$.}
\end{figure}

The shaded region in Fig.\,1 indicates the forbidden values of
the magnetospheric radius that exceed the corotation radius,
$R_m>R_c$. At such $R_m$ a rapidly rotating neutron
starmagnetosphere would produce a centrifugal barrier for the
accreting matter, inhibiting its penetration inward --- the
propeller regime would be switched on. Previously, it has
already been mentioned that for this reason, no efficient
accretion is possible at $\dot{M}_0<\dot{M}_{mc}$. It can be
seen from Fig.\,1 that in the case of a strong magnetic field
$B_*$, the situation when no direct accretion onto the neutron
star is possible at any $\dot{M}_0$ is realistic. Figure\,2
shows that this is actually the case. The solid line in this
figure indicates the accretion rate $\dot{M}_{mc}$ at which the
magnetospheric radius $R_m$ is equal to the corotation radius
$R_c$ (Eq. \ref{mdmc}) as a function of the magnetic field
strength $B_*$. In the shaded region to the right of this line
$R_m\ge R_c$; therefore, no accretion is possible here due to
the propeller effect. The dashed line in this figure indicates
the accretion rate $\dot{M}_{ms}$ at which the magnetospheric
radius $R_m$ is equal to the radius $R_s$ (Eq. \ref{mdms}) as a
function of $B_*$.  Above this curve $R_m<R_s$. As has already
been said, accretion flow spherization, a strong outflow of
matter under radiation pressure, and a drop in luminosity to
$\simeq 2 L_{\rm ed}$ begin here. Direct (efficient) accretion
is possible only in the $\dot{M}_0-B_*$ region lying between
these lines, to the left of their intersection. The limiting
field at which direct accretion is still possible corresponds to
the point of their intersection:
\begin{equation}
B_*^{\rm max}\simeq 4.4\times10^{13}\ R_{12}^{-3} m_* p_*^{3/2}\  \mbox{G}.
\end{equation}
However, it should be noted that at high accretion rates, when
$R_m<R_s$, being in the shaded region in comparison with being
outside it makes very little difference observationally --- as
before, we will see a source with a nearly Eddington total
luminosity. This luminosity will be emitted by the accretion
disk at large, $R>R_s$ , distances from the neutron star and the
outflowing envelope in the region $R_m^{\rm
  low}<R<R_s$. Obviously, the accreting matter does not fall
below the radius $R_m^{\rm low},$ so that it is impossible to
record any radiation pulsations in this regime of accretion.
Given what has been said above about the softness of the
radiation spectrum for such a source, it will most likely be
impossible to determine whether the two-fold drop in its
luminosity is associated with the excess of $R_m$ above $R_c$ or
an excessively narrow and hard range of its observations.

\section*{THE NEUTRON STAR SPINUP RATE}
\noindent
Although the measured long-term spinup rate of the neutron star
in the three discovered ULX pulsars,
$\dot{\nu}=-\dot{P}_sP_s^{-2}\sim (1-6)\times10^{-10}\ \mbox{Hz
  s}^{-1},$ exceeds the spinup rate of the neutron star in
ordinary X-ray pulsars by an order of magnitude or more (see the
table), it turns out to be several times lower than the spinup
rate expected for these sources, given the observed essentially
super-Eddington accretion rate,
\begin{equation}\label{dotnu}
\dot{\nu}=(GM_* R_m)^{1/2}\frac{\dot{M}_0}{2\pi I} \simeq
1.4\times10^{-9}\ \dot{m}_{20}^{6/7} m_*^{3/7} \mu_3^{2/7}
I_{45}^{-1}\ \mbox{Hz s}^{-1}.
\end{equation}
Here, $I_{45}$ is the moment of inertia $I _*$ of the neutron
star (normalized to its standard value of $10^{45}\ \mbox{g
  cm}^{2}$).  Such a slow spinup is naturally explained in the
scenario of supercritical accretion onto these pulsars proposed
above. Indeed, the estimate (\ref{dotnu}) refers only to the
high luminosity state of these sources. During their low state
the actual accretion rate near the magnetosphere decreases
considerably to $\dot{M}_0\,R_m^{\rm low}/R_s
\simeq\dot{M}_{ms};$ the spinup rate of the neutron star drops
accordingly.  Moreover, it should be noted that in this state
there is not disk accretion, which is capable of efficiently
transferring the angular momentum of Keplerian motion to the
neutron star, but almost quasi-spherical accretion of matter
that lost much of its angular momentum. The angular momentum is
transferred only in the narrow circular region which width is
much smaller than the real width of the accretion disk. The
foregoing implies that an efficient spinup of the neutron star
in the ULX pulsars occurs only during a certain fraction of the
entire time of their active existence, while in the remaining
time they barely spin up. For this reason, the mean spinup
determined from long time intervals turns out to be appreciably
smaller than their maximum spinup during the episodes of direct
super-Eddington accretion.

\section*{CONCLUSIONS}
\noindent
We gave an explanation for the bimodality of the X-ray
luminosity distribution of ULX pulsars. The transition from the
high to low state of these sources was explained by accretion
flow spherization when a certain accretion rate is exceeded. In
this case, the luminosity of the source drops to a nearly
Eddington level of $(1-2) L_{\rm ed}.$ The observed X-ray
luminosity can be even lower, given the softness of the
radiation spectrum forming in the envelope of matter outflowing
from the accretion disk due to the radiation pressure. Apart
from the rate of change of the accretion rate, the transition
rate between the states is determined by the time it takes for
the neutron star magnetosphere to be rearranged, the speed of
the mass transfer through the disk, and the outflow velocity of
the excess of matter.

The accretion-driven spinup rate of the neutron star in the low
state decreases considerably compared to the spinup rate in the
high state. This allows the mean, insufficiently high measured
spinup rate of the ULX pulsars, lower than the expected one by
several times at given accretion rates, to be explained.\\

\section*{ACKNOWLEDGMENTS}
\noindent
This work was financially supported by the Program of the
President of the Russian Federation for support of leading
scientific Schools (grant NSh-10222.2016.2) and the
``Transitional and Explosive Processes in Astrophysics''
Subprogram of the Basic Research Program P-7 of the Presidium of
the Russian Academy of Sciences.

\pagebreak   

\hfill {\it Translated by V. Astakhov\/}

\begin{references}

\reference{M. Bachetti, F. A. Harrison, D. J. Walton, B. W. Grefenstette,
D. Chakrabarty, F. F\"{u}rst, D. Barret, A. Beloborodov, et al.},
\nat\ {\bf 514}, 202 (2014).

\reference{M. M. Basko and R. A. Sunyaev}, 
\aap\ {\bf 42}, 311 (1975).

\reference{M. M. Basko and R. A. Sunyaev}, 
\mnras\ {\bf 175}, 395 (1976).

\reference{S. Campana, F. Brivio, N. Degenaar, S. Mereghetti, R. Wijnands,
P. D'Avanzo, G.L. Israel, and L. Stella}, 
\mnras\ {\bf 441}, 1984 (2014).

\reference{S. Campana, L. Stella, G.L. Israel, A. Moretti,
  A.N. Parmar, and M. Orlandini},
\apj\ {\bf 580}, 389 (2002).

\reference{S. Campana, L. Stella, and J.A. Kennea},
\apj\ {\bf  684},  L99 (2008).

\reference{W.-C. Chen},
\mnras\ {\bf 465},  L6 (2017).

\reference{R. H. D. Corbet},
\mnras\ {\bf 220}, 1047 (1986).

\reference{R. H. D. Corbet},
\apj\ {\bf 457}, L31 (1996).

\reference{K. Davidson and J. P. Ostriker}, 
\apj\ {\bf 179}, 585 (1973).

\reference{K. Y. Eksi, I. C. Andac, S. Cikintoglu,
  A. A. Gencali, C. G\"{u}ng\"{o}r, and F. \"{O}ztekin},
\mnras\ {\bf 448}, L40 (2015).

\reference{P. Ghosh and F. K. Lamb},
\apj\ {\bf 223}, L83 (1978).

\reference{S. A. Grebenev},
Proceedings of Science, {\bf 96}, 60 (Proc. of the
Conference ``The Extreme Sky: Sampling the Universe above 10
keV'', Otranto, Italy, October 13--17, 2009). 

\reference{S. A. Grebenev and R. A. Sunyaev},
\astl\ {\bf 33}, 149 (2007).

\reference{E. P. J. van den Heuvel},
J. Astrophys. Astron. {\bf 5}, 209 (1984).

\reference{A. F. Illarionov and R. A. Sunyaev},
\aap\ {\bf 39}, 185 (1975).

\reference{G. L. Israel, A. Belfiore, L. Stella, P. Esposito, P. Casella,
A. De Luca, M. Marelli, A. Papitto, et al.}, 
Science  {\bf 355}, 817; arXiv:1609.07375v1 (2017a).

\reference{G. L. Israel, A. Papitto, P. Esposito, L. Stella, L. Zampieri,
A. Belfiore, G.A. Rodriguez Castillo, A. De Luca, et al.}, 
\mnras\  {\bf 466}, L48;  arXiv:1609.06538v1 (2017b).

\reference{W. Kluzniak and J.-P. Lasota},
\mnras\ {\bf 448}, L43 (2015).

\reference{W. H. G. Lewin, J. van Paradijs, and R. E. Taam},
\ssr\ {\bf 62}, 223 (1993).

\reference{V. M. Lipunov},
\sva\ {\bf 26}, 54 (1982).

\reference{G. V. Lipunova},
\astl\ {\bf 25}, 508 (1999).

\reference{Yu. E. Lyubarskii and R. A. Sunayev},
\sval\ 14, 390 (1988).

\reference{M. Lyutikov},
arXiv:1410.8745 (2014).

\reference{A. A. Mushtukov, V. F. Suleimanov, S. S. Tsygankov,
  and J. Poutanen},
\mnras\ {\bf 454}, 2539 (2015).

\reference{S. dall'Osso, R. Perna, and L. Stella}, 
\mnras\ {\bf 449}, 2144 (2015).

\reference{B. Paczynski},
Acta Astronomica {\bf 42}, 145 (1992).

\reference{K. Postnov, L. Oskinova, and J.M. Torrej\'on},
\mnras\ {\bf 465}, L119 (2017).

\reference{N. I. Shakura and R. A. Sunyaev},
\aap\ {\bf 24}, 337 (1973).

\reference{Y. Shao and X.-D. Li},
\apj\ {\bf 802}, 131 (2015).

\reference{H. Tong}, 
Res. Astron. Astrophys.\ {\bf 15}, 517 (2015).

\reference{S. S. Tsygankov, A. A. Mushtukov, V. F. Suleimanov,
  and J. Poutanen},
\mnras\ {\bf  457}, 1101 (2016a).

\reference{S. S. Tsygankov, A. A. Lutovinov, V. Doroshenko, A. A. Mushtukov,
V. Suleimanov, and J. Poutanen},
\aap\ {\bf 593}, A16 (2016b).

\end{references}
\end{document}